\documentclass[12pt]{article}
\usepackage{hyperref}
\usepackage{graphicx}
\usepackage{amsmath}
\usepackage{amssymb}
\usepackage[numbers,compress]{natbib}
\usepackage{amsmath}

\begin{document}

\title{Light from Schwarzschild black holes in de Sitter expanding universe}

\author{Ion I. Cot\u aescu\\ {\small \it  West University of Timi\c soara,}\\
  {\small \it V. P\^ arvan Ave. 4, RO-1900 Timi\c soara, Romania}\\
 {\small \it i.cotaescu@e-uvt.ro}}

\maketitle
\begin{abstract}
A new  method is applied for deriving the redshift and shadow of a Schwarzschild black hole moving freely in the de Sitter expanding universe as recorded by a co-moving remote observer. This method is manly algebraic  focusing on the transformation of the conserved quantities under the de Sitter isometry relating the black hole proper co-moving frame to observer's one. Hereby one extracts the general expressions of the redshifts  and shadows of the black holes having peculiar velocities but their expressions are too extended to be written down here. Therefore, only some particular cases and intuitive expansions are presented while the complete results are given in an algebraic code on computer \cite{Comp}.

Pacs: 04.02.Cv and  04.02
\end{abstract}


 \newpage
\section{Introduction}
%\label{intro}

The  light emitted by cosmic objects is one of the principal sources of empirical data in astrophysics. An  important accessible observable is the redshift which encapsulates information about the cosmic expansion and possible peculiar velocity of the observed object.  For separating these two contributions one combined so far the Lema\^ itre rule \cite{L1,L2} of Hubble's law \cite{Hubb}, governing the cosmological effect, \cite{SW,H0,H} with the usual theory of the Doppler effect of special relativity   
\cite{LL} even though there are evidences that our universe is expanding. 

Recently we proposed an improvement of this approach replacing the special relativity with our de Sitter relativity \cite{CdSR1,CdSR2} where local charts of the same type, playing the role of inertial frames, are related among themselves through de Sitter isometries. In the case of the longitudinal Doppler effect, when a point-like source is moving along the axis observer-source, we obtained a rdshift formula having a new term combining the cosmological and kinetic contributions in a non-trivial manner \cite{CDop}.

The next step is to extend this method to the black holes which, in general,  are no  point-like sources. The light emitted by a black hole comes from an apparent source situated on a sphere, surrounding the black hole, which is observed as the black hole shadow. When this is not negligible, as in the case of  the object M87 \cite{BH1,BH2},  and the black hole may have a peculiar velocity, the Doppler effect is no longer longitudinal such that the transverse contributions due to the black hole shadow  must be evaluated.  This can be done only by studying simultaneously the Doppler effect and black hole shadow in the same theoretical framework.  

The black hole shadow was extensively studied by many authors which used a geometrical method which exploits the tangent to the null geodesic in the point where this meets the observer. Thus the shadows of the non-rotating Schwarzschild black holes were studied in  Minkowski space-time \cite{Sy} or  in expanding universes \cite{SS1,SS2,SS01,SS02,SS03,SS3} while for the rotating black holes the Kerr \cite{K1,K2} or Kerr-de Sitter \cite{KdS1,KdS2,KdS3,KdS4} metrics  were considered.  Other recent studies focused on more complex models which were studied with the same geometric method \cite{M1,M2,M3,M4,M5,M6,M7,M8,M9,M10,M11,M12,M13,M14,M15,M16} but which is not suitable for studying the Doppler effect. 

As in this paper we would like to study simultaneously the redshifts and shadows of the Schwarzschild black holes having peculiar velocities in de Sitter expanding universe we give up the geometric method adopting the same algebraic approach as in Ref. \cite{CDop}. This is based on our de Sitter relativity we use now for relating the moving black hole proper frames to those of remote co-moving observers freely falling in the de Sitter expanding universe.   

We start supposing that our expanding  universe is satisfactory described by the expanding portion of a $(1+3)$-dimensional de Sitter manifold. As the actual observations show with reasonable accuracy that this universe is spatially flat, we consider only local charts with  Painlev\' e coordinates \cite{Pan} since these have flat space sections. In these local charts, we call from now {\em co-moving} frames, the coordinates we use are the {\em cosmic} time and Cartesian or spherical space coordinates.  These frames may carry observers  related among themselves through the de Sitter isometries which transform simultaneously the coordinates and the conserved quantities \cite{CdSR1,CdSR2}. 

The Schwarzschild black holes in de Sitter expanding universe are usually considered in proper frames with static coordinates and Kottler metric \cite{Kot}, known mainly as the Schwarzschild-de Sitter proper frames. However, here we give up the static frames preferring the corresponding co-moving frames with Painlev\' e coordinates whose metrics have the same asymptotic behaviour as the metric of the observer co-moving frame.  Then for a remote observer the black hole co-moving frame  appears as an empty de Sitter one  which can be related to observer's frame through a de Sitter isometry, in accordance with the relative motion of the black hole with respect to observer. 

We assume that, at the initial moment when the photon is emitted by black hole, this is translated and has a relative velocity with respect to the remote observer.  Then a suitable isometry will give the conserved quantities measured in observer's frame we need for extracting physical results without resorting to geodesics or other geometric quantities. For this reason we say that our method is {\em algebraic}. We use this method for deriving the redshift and shadow of a non-rotating Schwarzschild black hole freely moving in the de Sitter expanding universe.

We start in the second section with a brief review of the metrics of the black hole and observer co-moving frames with Cartesian or spherical coordinates,  revisiting the equation giving the geodesic shapes in the black hole co-moving frames. Starting with this equation we present in the next section the null geodesics around the black hole determining its shadow. The fourth section is devoted to the de Sitter isometries relating the conserved quantities measured in different co-moving frames.  In the next section we obtain our new results assuming that for a remote observer the light is emitted by an apparent source on a rectilinear de Sitter geodesic whose conserved quantities can be determined. Furthermore, by using an isometry formed by a translation followed by a Lorenzian isometry we obtain the conserved quantities in observer's co-moving frame from which we extract the observed redshift and shadow of the moving black hole.  More specific, the redshift results from the observed energy while the angular radius of the black hole shadow is derived from the photon angular momentum which must vanish when the photon meets the origin of observer's co-moving frame.  These results are elementary but with a large number of terms that cannot be written here in the general case of a moving black hole. Consequently, we restrict ourselves to present here  their series expansions with respect to a small parameter, the particular case when the relative velocity vanishes and the flat limit. The complete results which cannot be written down here are given in an algebraic code on computer  \cite{Comp}.   In order to convince oneself that the flat limit is correct we derive in  App. B  the results that can be obtained by applying our algebraic method to a Schwarzschild black hole in Minkowski flat space-time.  Finally we present some concluding remarks.  

As our approach may be applied even in quantum theory we introduce a special notation denoting by $\omega_H=\sqrt{\frac{\Lambda}{3}}c$ the de Sitter Hubble constant (frequency) since $H$ is reserved for the Hamiltonian operator \cite{CGRG}. Moreover,  the Hubble time  $t_H=\frac{1}{\omega_H}$ and the Hubble length   $l_H=\frac{c}{\omega_H}$ will have the same form  in the natural Planck units with $c=\hbar=G=1$ we use here.

\section{Co-moving frames}

The  frames of the spherically symmetric static systems in a $(1+3)$-dimensional isotropic pseudo-Riemannian manifold, $(M,g)$, are static local charts $\{x\}$ with spherical symmetry whose coordinates $x^{\mu}$ ($\alpha,\mu,\nu,...=0,1,2,3$) can be chosen in different manners.  In what follows we consider either Cartesian space coordinates ${\bf x}=(x^1,x^2,x^3)$ or associated spherical ones $(r,\theta,\phi)$ with $r=|{\bf x}|$. The traditional {\em static} frames, $\{t_s,{\bf x}\}$ or $\{t_s,r,\theta,\phi\}$, depend on  the static time  $t_s$ having  the line elements
\begin{eqnarray}
ds^2&=&f(|{\bf x}|)dt_s^2+\left(\frac{{\bf x}\cdot d{\bf x}}{|{\bf x}|}\right)^2\left(1-\frac{1}{f(|{\bf x}|)}\right)-d{\bf x}\cdot d{\bf x}\label{s1c}\\
&=&f(r)\, dt_s^2-\frac{dr^2}{f(r)}-r^2 d\Omega^2\,,\label{s1s}
\end{eqnarray}
where $d\Omega^2=d\theta^2+\sin^2\theta\, d\phi^2$. These line elements can be put at any time in Painlev\' e forms \cite{Pan}, 
\begin{eqnarray}
ds^2&=&f(|{\bf x}|)dt^2+\frac{2}{|{\bf x}|}\sqrt{1-f(|{\bf x}|)}\, {\bf x}\cdot d{\bf x}\,dt-d{\bf x}\cdot d{\bf x}\label{sc}\\
&=&f(r)dt^2+2\sqrt{1-f(r)}\,dtdr-dr^2-r^2 d\Omega^2\,,\label{ss}
\end{eqnarray}
substituting in Eqs. (\ref{s1c}) and (\ref{s1s}) 
\begin{equation}\label{subs1}
t_s=t+\int dr \frac{\sqrt{1-f(r)}}{f(r)}\,,
\end{equation}
where $t$ represents the {\em cosmic} time of the frames having flat space sections with Cartesian, $\{t, {\bf x}\}$, or spherical, $\{t, r,\theta,\phi\}$, coordinates.  

Here we focus on a Schwarzschild  black hole of mass $M$ embedded in the de Sitter expanding universe for which the metric (\ref{s1s}) of its static frame has the Kottler  \cite{Kot} (or Schwarzschild-de Sitter) form with
\begin{equation}\label{frM}
f(r)=1-\frac{2M}{r}-\omega_H^2 r^2 \,.
\end{equation} 
where, as mentioned before,  $\omega_H$ is the de Sitter Hubble constant in our notation.  The corresponding frames with Painlev\' e coordinates have the asymptotic behaviour of  the de Sitter co-moving  frames with  $f(r) \to f_0(r)=1-\omega_H^2 r^2$. For this reason we say that  the black hole proper frames with Painlev\' e coordinates,  denoted by $\{t,{\bf x}\}_{BH}$ and  $\{t,r,\theta,\phi\}_{BH}$, are the {\em co-moving}  proper frames of the Schwaezschild black hole in de Sitter expanding universe. 

The proper frames of the remote observers, $\{t,{\bf x}\}$ and  $\{t,r,\theta,\phi\}$, located in the asymptotic zone, are genuine de Sitter co-moving frames where the astronomical observations are performed and recorded.  The observers stay at rest in  origins of their own proper frames evolving along the unique time-like Killing vector field of the de Sitter geometry which  is not time-like everywhere but has this property just in the null cone where the observations are allowed \cite{CGRG}.   

Here we use simultaneously Cartesian and spherical coordinates since 
the Cartesian coordinates are  suitable for studying the conserved quantities and the transformation rules under isometries while the spherical coordinates help one to integrate the geodesic equations. For example, the spherical symmetry is obvious in Cartesian coordinates where the metrics (\ref{s1c}) and (\ref{sc}) are invariant under the global rotations  $x^i\to R^i_jx^j$ such that we can use the vector notation.  On the other hand, only in spherical coordinates one can integrate the geodesics equations in the black hole co-moving frame we revisit briefly in the next. 

In the frame  $\{t,r,\theta,\phi\}_{BH}$ with the line element (\ref{ss}) the conserved quantities along geodesics are the energy $E$ and angular momentum ${\bf L}$. These give rise to the prime integrals of  a geodesic of  a particle of mass $m$ moving  in the equatorial plane of the black hole (with fixed  $\theta=\frac{\pi}{2}$) as
\begin{eqnarray}
E&=&f(r)\dot t +\sqrt{1-f(r)}\,\dot r\,,\label{E}
\\
L&=&r^2 \dot \phi \,,\label{L}
\end{eqnarray}
where  'dot' denotes the derivatives with respect to the affine parameter $\lambda$ which satisfies  $ds=m\,d\lambda$. The third prime integral comes from the line element in the equatorial plane that reads,
\begin{equation}\label{spp2}
f(r){\dot t}^2+2\sqrt{1-f(r)}\,\dot t\dot r-{\dot r}^2-r^2 {\dot \phi}^2=m^2\,.
\end{equation}
as it results from Eq. (\ref{ss}). Hereby one may derive the  function $r(\phi)$ substituting 
\begin{equation}\label{Bin1}
r\to r(\phi)\,, \quad \dot r \to\frac{dr(\phi)}{d\phi}\dot\phi=\frac{L}{r(\phi)^2}\frac{dr(\phi)}{d\phi}\,.
\end{equation}
After a little calculation, combining the above prime integrals, one obtains the well-known equation 
\begin{equation}\label{Bin}
\left(\frac{dr(\phi)}{d\phi}\right)^2-r(\phi)^4\frac{E^2}{L^2}+r(\phi)^2f[r(\phi)] 
\left(1+r(\phi)^2\frac{m^2}{L^2}\right)=0\,,
\end{equation} 
giving the geodesic shapes but which is not enough for finding the time behaviour of the functions $\phi(t)$ and $r(t)=r[\phi(t)]$ for which one must apply special methods \cite{Gib}. 

Note that Eq. (\ref{Bin}) derived in the co-moving frame is the same as that of the static frame since this equation is static giving only the shape of trajectory in the same space coordinates.  In fact, the time evolution on geodesics is quite different in the static and co-moving frames.   

\section{Light around black holes}

The problem of the gravitational lensing which has a long history \cite{El1} was studied in general relativity first by Einstein and Eddington \cite{Ed} but was solved by Darwin \cite{D1,D2} which derived the null geodesics around the photon sphere  of a Schwarzschild black hole in the flat Minkowski space-time. Applying the same  commonly used method \cite{V1,V2,V3,V4,V5,V6} we may inspect briefly the null geodesics in the co-moving frame $\{t,r,\theta,\phi\}_{BH}$ of the Schwarzschild-de Sitter system.

{ \begin{figure}
  \centering
    \includegraphics[scale=0.55]{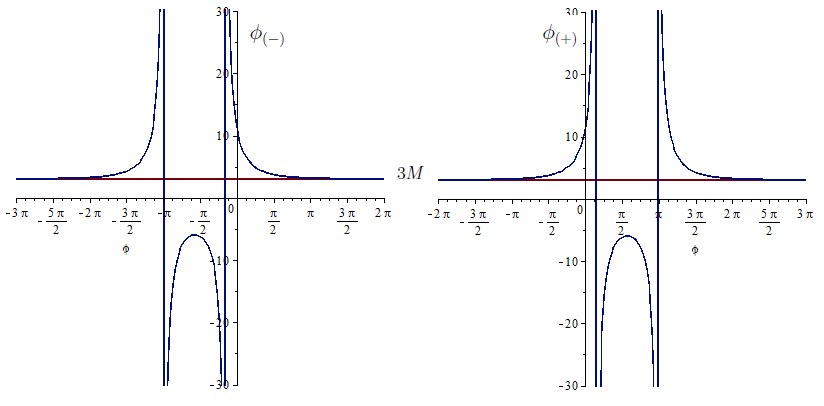}
    \caption{The functions $r_{(\pm)}(\phi)$ of the spiral photon geodesics closest to the  photon sphere of radius $3 M$.}
  \end{figure}}

The shapes of the photon geodesics are given by the functions $r(\phi)$ which satisfy Eq.  (\ref{Bin}) with $m=0$ that now reads
\begin{equation}\label{Bin1}
\left(\frac{dr(\phi)}{d\phi}\right)^2-r(\phi)^4\frac{E_{ph}^2}{L_{ph}^2}+r(\phi)^2f[r(\phi)] 
=0\,,
\end{equation} 
This equation has two types of solutions, namely circular geodesics on the photon sphere and associated spiral geodesics \cite{D1}. The circular geodesics satisfy simultaneously the conditions 
\begin{equation}
\frac{dr(\phi)}{d\phi}=\frac{d^2 r(\phi)}{d\phi^2}=0\,,
\end{equation}
giving the radius of the photon sphere  $r_{ph}=3M$ and the mandatory condition 
\begin{equation}\label{Lps}
L_{ph}=\pm\frac{3\sqrt{3}M E_{ph}}{\sqrt{1-27 \omega_H^2 M^2}}\,,
\end{equation}
derived in Ref. \cite{SS1}. Thus the photons with circular  geodesics are trapped on the photon sphere without escaping outside. Furthermore, by substituting the condition (\ref{Lps}) in Eq. (\ref{Bin1}) we obtain the equation
\begin{equation}\label{eqph}
27 M^2\left(\frac{dr(\phi)}{d\phi}\right)^2-54 M^3 r(\phi)+27 M^2 r(\phi)^2-r(\phi)^4=0\,,
\end{equation} 
which is independent on the Hubble de Sitter constant $\omega_H$. Apart from the circular geodesics, this equation allows the solutions 
\begin{equation}\label{rphi}
r_{(\pm)}(\phi)=3M\frac{\left(6M e^{\mp \phi}+1\right)^2}{\left(6M e^{\mp\phi}+1\right)^2-36 M e^{\mp\phi}}\,,
\end{equation}
known as the spiral geodesics \cite{D1}. These are determined up to a rotation,  $\phi\to \phi-\phi_0$, fixing the origin of this angular coordinate. For example, if we translate the arguments of the functions $r_{(\pm)}$ as $\phi\to\phi^{\pm}=\phi\pm \ln 6M$ then we recover the elegant Darwin form \cite{D1}
\begin{equation}
\frac{1}{r_{(\pm)}(\phi^{\pm})}=-\frac{1}{6M}+\frac{1}{2M}\, \tanh^2 \left(\frac{\phi^{\pm}}{2}\right)\,,
\end{equation}
of the spiral geodesics. Thus we may conclude that the presence of the de Sitter gravity is encapsulated only in Eq. (\ref{Lps}) while the photon sphere and the shapes of the spiral geodesics remain the same as in Minkowski  flat space-time (when $\omega_H=0$).

{ \begin{figure}
  \centering
    \includegraphics[scale=0.65]{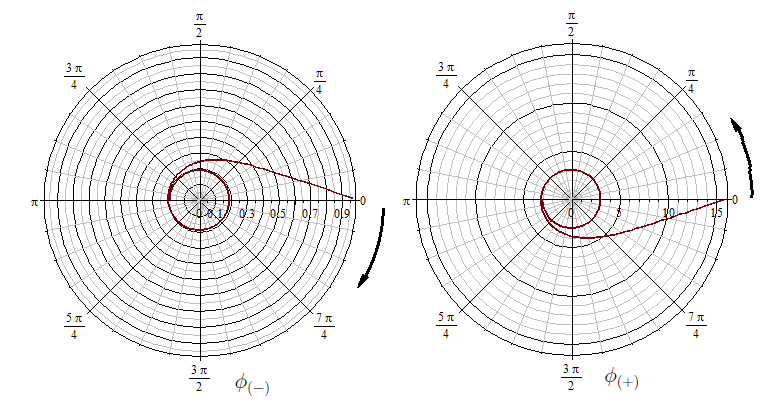}
    \caption{The functions $r_{(\pm)}(\phi)$ of the spiral photon geodesics rolling out and escaping from the photon sphere of radius $3 M$.}
  \end{figure}} 
  
The spiral geodesics  are symmetric in the sense that $r_{(\pm)}(\phi)=r_{(\mp)}(-\phi)$,  having vertical asymptotes at 
\begin{equation}\label{ang}
\phi_{(+)\,1,2}=- \ln \frac{2\mp\sqrt{3}}{6M}\,,\quad \phi_{(-)\,1,2}= \ln \frac{2\mp\sqrt{3}}{6M}\,,
\end{equation}
In  Fig. 1. we see that the functions $r_{(\pm)}(\phi)$  are defined on the domains $(-\infty, \phi_{(\pm)\, 1})\cup (\phi_{(\pm)\,2 },\infty)$  since between the vertical asymptotes their values are negative having no physical meaning. It is interesting that this opaque window is independent on the black hole mass, 
\begin{equation}
|\phi_{(\pm)\,2}-\phi_{(\pm)\, 1}|=\ln \frac{2+\sqrt{3}}{2-\sqrt{3}}\simeq0.8384\, \pi\,.
\end{equation}
On the physical domain these trajectories remain outside the photon sphere, $r_{(\pm)}(\phi)>3M$, but approaching to this for large $|\phi|$ since   
\begin{equation}
\lim_{|\phi|\to \infty}r_{(\pm)}(\phi)=3M. 
\end{equation}
This gives us the image of the spiral geodesics rolled out around the photon sphere  (as in Fig. 2.)  escaping outside only when $\phi$ is approaching to the values  (\ref{ang}) where the functions $r_{(\pm)}$  can take larger values near singularities.

The geodesics $r_{(\pm)}(\phi)$ are the closest trajectories to the photon sphere of the first photons that can be observed at the limit of the black hole shadow. Therefore, for studying this shadow and the associated  redshift we have to consider only these photons.  

\section{de Sitter isometries}  

The  de Sitter co-moving frames play the role of inertial frames being related among themselves through de Sitter isometries as in our de Sitter relativity \cite{CdSR1,CdSR2}.    Moreover, the system black  hole-observers has the asymptotic de Sitter symmetry which governs the relative motion of the black hole with respect to different observers such that we may use these isometries for relating the observers co-moving frames to the black hole one. 

The de Sitter isometries can be studied easily since this manifold is a hyperboloid  of radius $1/\omega_H$ embedded  in the five-dimensional flat space-time $(M^5,\eta^5)$ of coordinates $z^A$  (labelled by the indices $A,\,B,...= 0,1,2,3,4$) and metric $\eta^5={\rm diag}(1,-1,-1,-1,-1)$. 
Any  local charts  can be introduced  giving the set of functions $z^A(x)$ which solve the hyperboloid equation,
\begin{equation}\label{hip}
\eta^5_{AB}z^A(x) z^B(x)=-\frac{1}{\omega_H^2}\,,
\end{equation}
giving the line element 
\begin{equation}\label{s}
ds^{2}=\eta^5_{AB}dz^A(x)dz^B(x)=g_{\mu\nu}dx^{\mu}dx^{\nu}\,.
\end{equation} 
The functions that introduce our Painlev\' e  coordinates are
\begin{eqnarray}
z^0(x)&=&\frac{1}{2\omega_H}\left[e^{\omega_H t}-e^{-\omega_H t}(1 - \omega_H^2{\bf x}^2)\right]\,,
\nonumber\\
z^i(x)&=&x^i \,, \label{Zx}\\
z^4(x)&=&\frac{1}{2\omega_H}\left[e^{\omega_H t}+e^{-\omega_H t}(1 - \omega_H^2{\bf x}^2)\right]\,.
\nonumber
\end{eqnarray}

The de Sitter isometry group is just the stable group $SO(1,4)$ of the embedding manifold $(M^5,\eta^5)$  that leave  invariant its metric and implicitly Eq. (\ref{hip}). Therefore, given a system of coordinates defined by the functions $z=z(x)$, each transformation ${\frak g}\in SO(1,4)$ gives rise to the isometry $x'\to x=\phi_{\frak g}(x')$ derived from the system 
\begin{equation}\label{iso}
z[\phi_{\frak g}(x')]={\frak g}z(x')\,.
\end{equation}
The  local charts related through these isometries play the same role as the  inertial frames of special relativity. 

The classical conserved quantities under de Sitter isometries  are given by  the  Killing vectors  $k_{(AB)}$ of the de Sitter manifold  \cite{CGRG} that are related to those of  $(M^5,\eta^5)$ as 
\begin{equation}
K^{(AB)}_Cdz^C=z^Adz^B-z^Bdz^A=k^{(AB)}_{\mu}dx^{\mu}\,,
\end{equation}
allowing us to derive the covariant components  of the Killing vectors in an arbitrary chart $\{x\}$ of  the de Sitter space-time as
\begin{equation}\label{KIL}
k_{(AB)\,\mu}=\eta^5_{AC}\eta^5_{BD}k^{(CD)}_{\mu}= z_A\partial_{\mu}z_B-z_B\partial_{\mu}z_A\,, 
\end{equation}
where $z_A=\eta_{AB}z^B$. The conserved quantities along the time-like geodesic of a particle of mass $m$  have the general form  ${\cal K}_{(AB)}(x,{\bf P})=\omega_H k_{(AB)\,\mu}\dot x^{\mu}$.  The conserved quantities with physical meaning are the  energy $E$, momentum ${\bf P}$, angular momentum ${\bf L}$ and a specific vector ${\bf Q}$, we called the adjoint momentum \cite{CGRG}. A geodesic in the  co-moving frame $\{t,{\bf x}\}$ \cite{CdSG}, 
\begin{equation}\label{geodS}
{\bf x}(t) ={\bf x}_0e^{\omega_H( t-t_0)}+\frac{{\bf P}\,e^{\omega_H t}}{\omega_H P^2}\left(\sqrt{m^2+P^2e^{-2\omega_H t_0}}-\sqrt{m^2 +P^2e^{-2\omega_H t}}\right)\,,
\end{equation}
depends only on the momentum ${\bf P}$ ($P=|{\bf P}|$) and the initial condition ${\bf x}(t_0)={\bf x}_0$ fixed at the time $t_0$. The conserved quantities  in an arbitrary point $(t,{\bf x}(t))$  of this geodesic read  \cite{CdSR1,CdSG},
\begin{eqnarray}
E&=&\omega_H\, {\bf x}(t)\cdot {\bf P}\,e^{-\omega_H t}+\sqrt{ m^2+{P}^{2}e^{-2\omega_H t}}\,,\label{Ene}\\
{\bf L}&=& {\bf x}(t)\land {\bf P}\,e^{-\omega_H t}\,,\label{La}\\
{\bf Q}&=&2\omega_H\, {\bf x}(t)E e^{-\omega_H t}+{\bf P}  e^{-2\omega_H t}[1-\omega_H^2{\bf x}(t)^2]\,.\label{Qa}\end{eqnarray}
satisfying the  obvious identity
\begin{equation}\label{disp}
E^2-\omega_H^2 {{\bf L}}^2-{\bf P}\cdot {\bf Q}=m^2
\end{equation}
corresponding to the first Casimir invariant of the $so(1,4)$ algebra \cite{CGRG}. In the flat limit, when $\omega_H\to 0$,  we have ${\bf Q} \to {\bf P}$ such that this identity  becomes just  the usual mass-shell condition, $E^2-{\bf P}^2=m^2$, of special relativity.   

The conserved quantities $E$, ${\bf P}$ and the new ones,
\begin{equation}\label{KR}
{\bf K}=-\frac{1}{2\omega_H}\left({\bf P}-{\bf Q}\right)\,, \quad {\bf R}=-\frac{1}{2\omega_H}\left({\bf P}+{\bf Q}\right)\,,
\end{equation}
form a skew-symmetric tensor on $M^5$,  
 \begin{equation}
{\cal K}(x,{\bf P})=
\left(
\begin{array}{ccccc}
0&\omega_H K_1&\omega_H K_2&\omega_H K_3&E\\
-\omega_H K_1&0&\omega_H L_3&-\omega_H L_2&\omega_H R_1\\
-\omega_H K_2&-\omega_H L_3&0&\omega_H L_1&\omega_H R_2\\
-\omega_H K_3&\omega_H L_2&-\omega_H L_1&0&\omega_H R_3\\
-E&-\omega_H R_1&-\omega_H R_2&-\omega_H R_3&0
\end{array}\right)\,,\label{KK}
\end{equation}
whose components  transform under the isometries $x'\to x=\phi_{\frak g}(x')$ defined by Eq. (\ref{iso}) as 
\begin{equation}\label{Kg}
{\cal K}(t,{\bf x},{\bf P})=\overline{\frak g}\,{\cal K}'(t',{\bf x}',{\bf P}')\overline{\frak g}^T\,,
\end{equation} 
where $\overline{\frak g}=\eta^5\,{\frak g}\,\eta^5$ \cite{CdSR1}.

Summarizing, we can say that the de Sitter isometries are generated globally by the $SO(1,4)$ transformations which determine the transformations of the coordinates and conserved quantities.  We have thus a specific relativity on the de Sitter space-time allowing us to study different relativistic effects in the presence of the de Sitter gravity. In what follows we use the Lorentzian isometries  defined in Ref. \cite{CdSR1} and  the translations presented in the App. A. 

\section{Observing light from black holes}

Le us consider now a mobile black hole in its proper co-moving frame $\{t',{\bf x}'\}_{BH}$ with the origin in $O_{BH}$ and a fixed remote observer in his own  co-moving frame $\{t,{\bf x}\}$ having the origin in $O$. We consider that the space Cartesian axes of these frames remain parallel  with the basis of unit vectors $({\bf e}_1,{\bf e}_2,{\bf e}_3)$ such that the geodesic of the emitted photon is in the plane $({\bf e}_1,{\bf e}_2)$.  In this geometry we assume that the photon is emitted at the initial moment $t=t'=0$ when the origin $O_{BH}$  is translated with $d$ and has the relative velocity ${\bf V}={\bf e}_1 V$ with respect to $O$. Note that the velocity ${\bf V}=\frac{\bf P}{M}$ is conserved depending on the conserved momentum ${\bf P}$ of the black hole geodesic observed by $O$. 

{ \begin{figure}
  \centering
    \includegraphics[scale=0.65]{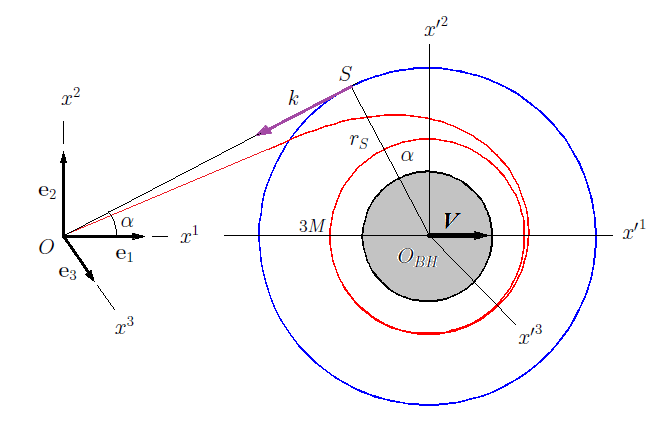}
    \caption{The geodesic $r_{(+)}(\phi)$ observed as a rectilinear de Sitter one of a photon of momentum $k$ emitted by the apparent source $S$ situated on the sphere of radius $r_S>3\sqrt{3} M$.}
  \end{figure}}

\subsection{Related conserved quantities}

A remote observer sees the photon of momentum ${\bf k}={\bf n}_k k$, energy $E_{ph}=k=|{\bf k}|$ and angular momentum  (\ref{Lps}), as emitted from an apparent source $S$ of position vector ${\bf n}_S r_S$  on the sphere of radius
\begin{equation}
r_S=\frac{|L_{ph}|}{k}=\frac{3\sqrt{3}M}{\sqrt{1-27 \omega_H^2 M^2}}\,,
\end{equation}   
which is the apparent radius of the black hole shadow. Therefore,  $r_S$ is the radius of the sphere hosting different photon sources $S$ that can be observed nearest to the black hole shadow (as in Fig. 3).  When  $\omega_H \to 0$ this becomes just the shadow radius  $3\sqrt{3}M$ derived in special relativity  \cite{D1}. 

The apparent trajectory of the emitted photon is a de Sitter null geodesic of momentum ${\bf k}$ that, according to Eq. (\ref{geodS}), reads 
\begin{equation}
{\bf x}_{ph}(t)={\bf n}_S r_S\,e^{\omega_H t}+{\bf n}_k\left(e^{\omega_H t}-1\right)\,,
\end{equation}
complying with the  initial condition ${\bf x}_{ph}(0)= {\bf n}_S r_S$. This geodesic depends on the orthogonal unit vectors that can be represented as    
\begin{eqnarray}
{\bf n}_k&=&-{\bf e}_1\cos\alpha - {\bf e}_2\sin\alpha\,,\\
{\bf n}_S&=& -{\bf e}_1\sin\alpha + {\bf e}_2\cos\alpha\,,
\end{eqnarray}
where the angle $\alpha$, giving the apparent direction of the photon, will depend on observer's position. We have thus  the opportunity  of defining  the de Sitter conserved quantities on this geodesic as in an apparent de Sitter empty co-moving frame  $\{t',{\bf x}'\}$  associated to $\{t',{\bf x}'\}_{BH}$. One vector is missing, namely the  adjoint momentum ${\bf Q}'$ that can be derived simply at the time $t=0$  according to Eq. (\ref{Qa}).  We complete thus the set of conserved quantities, 
\begin{eqnarray}
E'&=&k\,,\label{Q1}\\
{\bf P}'&=& {\bf n}_k\, k\,,\\
{\bf L}'&=&{\bf e}_3\, r_S k\,,\\
{\bf Q}'&=&{\bf n}_S\, 2\omega_H r_S k+ {\bf n}_k\, k\left(1-\omega_H^2 r_S^2\right)\,,\label{Q2}
\end{eqnarray}   
which satisfy the condition (\ref{disp}).

Now we can deduce how these conserved quantities  are measured by the fixed observer $O$ since the observer frame $\{t,{\bf x}\}$ and the apparent black hole one, $\{t',{\bf x}'\}$, are related through an isometry, $x=\phi_{\frak g}(x')$, of the de Sitter relativity \cite{CdSR1}.  According to our hypotheses, this is  generated by the SO(1,4) transformation,  
\begin{equation}\label{troc}
{\frak g}={\frak g}({\bf V}) {\frak g}({\bf a})\,,
\end{equation}
formed by a translation (\ref{tr1}) of parameter ${\bf a}={\bf e}_1d=(d,0,0)$, having the form \cite{Nach,CdSR1},  
\begin{equation}\label{tr}
{\frak g}({\bf a})=\left(
\begin{array}{ccccc}
1+\frac{1}{2}\,\omega_H^2 {d}^2&\omega_H d&0&0&\frac{1}{2}\,\omega_H^2 {d}^2\\
\omega_H d&1&0&0&\omega_H d\\
0&0&1&0&0\\
0&0&0&1&0\\
-\frac{1}{2}\,\omega_H^2 {d}^2&-\omega_H d&0&0&1-\frac{1}{2}\,\omega_H^2 {d}^2
\end{array}\right)\,,
\end{equation}
followed by the Lorentz boost   
\begin{equation}\label{Lor}
{\bf V}=(V,0,0)~\to~
{\frak g}({\bf V})=\left(
\begin{array}{ccccc}
\frac{1}{\sqrt{1-V^2}}&\frac{V}{\sqrt{1-V^2}}&0&0&0\\
\frac{V}{\sqrt{1-V^2}}&\frac{1}{\sqrt{1-V^2}}&0&0&0\\
0&0&1&0&0\\
0&0&0&1&0\\
0&0&0&0&1
\end{array}\right) \,,
\end{equation}
of the particular  Lorenzian isometry we need here \cite{CdSR1}. 

Applying then the transformation  (\ref{Kg}) with ${\frak g}$ given by Eq. (\ref{troc}) we obtain the conserved quantities observed by $O$.  This calculation is elementary but complicated involving many terms that can be manipulated only  by using suitable algebraic codes on computer. For presenting the final result it is convenient to introduce the notation
\begin{equation}
\delta=\omega_H d=\frac{d}{l_H} \,,\quad \xi=\frac{r_S}{d}\,,
\end{equation}
which allows us to write down the conserved quantities in observer's frame as
\begin{eqnarray}
E&=&\frac{ k}{\sqrt{1-V^2}}\left[ 1-V\cos \alpha +\delta(V-\cos\alpha -\xi V\sin\alpha ) \right.\nonumber\\
&&\hspace*{20mm}\left. -\frac{\delta^2}{2}(1-\xi^2)V\cos\alpha   \right]\,,\label{Efin}
\end{eqnarray}
\begin{eqnarray}
P^1&=&k\left(\frac{1}{\sqrt{1-V^2}}-1\right)\left[\delta(1-\xi\sin\alpha)-\frac{\delta^2}{2}(1-\xi^2)\cos\alpha\right]\nonumber  \\
&&\hspace*{20mm}+\frac{k}{\sqrt{1-V^2}}\left[V(1-\delta \cos\alpha)-\cos\alpha\right]\,,
\end{eqnarray}
\begin{eqnarray}
P^2&=& k\left(\frac{1}{\sqrt{1-V^2}}-1\right)\left[\delta\xi(\delta-\cos\alpha)-\frac{\delta^2}{2}(1+\xi^2)\sin\alpha \right] \nonumber\\
&&\hspace*{20mm}-k\sin\alpha+\frac{k\delta V}{\sqrt{1-V^2}}(\xi -\sin\alpha) \,,
\end{eqnarray}
\begin{eqnarray}
Q^1&=&k\left(\frac{1}{\sqrt{1-V^2}}+1\right)\left[\delta(1-\xi\sin\alpha)-\frac{\delta^2}{2}(1-\xi^2)\cos\alpha\right]\nonumber  \\
&&\hspace*{20mm}+\frac{k}{\sqrt{1-V^2}}\left[V(1-\delta \cos\alpha)-\cos\alpha\right]\,,
\end{eqnarray}
\begin{eqnarray}
Q^2&=& k\left(\frac{1}{\sqrt{1-V^2}}+1\right)\left[-\delta\xi(\delta-\cos\alpha)+\frac{\delta^2}{2}(1+\xi^2)\sin\alpha \right] \nonumber\\
&&\hspace*{20mm}-k\sin\alpha-\frac{k\delta V}{\sqrt{1-V^2}}(\xi -\sin\alpha) \,,
\end{eqnarray}
\begin{eqnarray}
L_3&=&\frac{k\delta}{\omega_H \sqrt{1-V^2}}\left[  \xi(1-V\cos\alpha)-\sin\alpha+\delta\xi V\right.\nonumber\\
&&\hspace*{20mm}\left.-\frac{\delta V}{2}(1+\xi^2)\sin\alpha \right]\label{Lfin}
\end{eqnarray}
while $L_1=L_2=P^3=Q^3=0$. As expected, these quantities satisfy the invariant identity (\ref{disp}). 

In other respects, we observe that all the vector components we derived above can take any real values in contrast with the energy which must remain positive definite. This condition is fulfilled only if
\begin{equation}\label{Vfin}
V<V_{lim}=\frac{2(1-\delta\cos\alpha)}{\delta^2(1-\xi^2)\cos\alpha-2\delta(1-\xi\sin\alpha)+2\cos\alpha}
\end{equation} 
This is in fact the mandatory condition for observing the photon in $O$ at finite time. When the relative velocity $V$ exceeds this limit then the photon cannot arrive in  $O$ at finite time because of the background expansion. Thus $V_{lim}$ defines a new velocity horizon restricting the velocities such that for very far sources with $\alpha=0$ and $\delta=\omega_H d\sim 1$ this limit vanishes.  

\subsection{Shadow and redshift}

The angle $\alpha$ which depends on the relative position between black hole and observer can be found simply imposing the condition $L_3=0$ when the photon trajectory is passing through the point $O$. Solving this equation for $L_3$ given by Eq. (\ref{Lfin}) we find
\begin{equation}
\tan \frac{\alpha}{2}=\frac{\delta V(1+\xi^2)+2 -\sqrt{\left[V\delta(1-\xi^2)+2\right]^2+4\xi^2(V^2-1)}}{2\xi(V\delta +V+1)}\,.
\end{equation}
Substituting then this angle in Eqs. (\ref{Efin}-\ref{Lfin}) and (\ref{Vfin}) we obtain as  final result all the conserved quantities measured by $O$ and the maximal velocity $V_{lim}$. 

Hereby we can extract the quantities of general interest namely $\sin\alpha$, measured in the black hole proper frame,  the angular radius 
\begin{equation}
\sin\alpha_{obs}=\frac{|P^2|}{P}\,,\quad P=\sqrt{{P^1}^2+{P^2}^2}\,,
\end{equation}
of the shadow, measured in observer's frame, and the redshift $z$ defined as 
\begin{equation}
1+z=\frac{E'}{E}=\frac{k}{E}\,.
\end{equation} 
Unfortunately, the exact expressions of these quantities have a huge number of terms that cannot be written here but can be manipulated on computer \cite{Comp} for  extracting significant particular cases or intuitive approximations.

The simplest particular case is when the black hole does not have a initial relative velocity with respect to $O$. Then by setting $V=0$ we obtain the simple formulas
\begin{eqnarray}
\sin\alpha_{obs}&=&\sin\alpha=\xi\,, \label{fina}\\
\frac{1}{1+z}&=&1-\omega_H d \sqrt{1-\xi^2}=1-\omega_H d\cos\alpha\,,\label{finz}
\end{eqnarray}
showing how the observations of the shadow and redshift are related each other.

In the general case of $V\not= 0$, we observe that the  expansions around $\xi=0$ are useful since this is the only parameter which remains very small in astronomical observations as long as the other ones have larger ranges, $0<\delta=\omega_H d<1$ and $0<V<V_{lim}$. Computing these series we write down here only the lowest terms 
\begin{eqnarray}
&&\sin\alpha= \frac{2(1+\omega_H d V-V)}{2+\omega_H d V}\,\xi+{\cal O}(\xi^3)\,, \label{finA0}\\
&&\sin\alpha_{obs}=\frac{2\xi\sqrt{1-V^2}}{2+\omega_H d V (1-\xi^2)}= \frac{2\sqrt{1-V^2}}{2+\omega_H d V}\,\xi +{\cal O}(\xi^3)\,,\label{finA}\\
&&\frac{1}{1+z}=\sqrt{\frac{1-V}{1+V}}\left(1-\omega_H d -\frac{\omega_H^2 d^2}{2}\,\frac{V}{1-V}\right) +{\cal O}(\xi^2)\,,\label{finZ}
\end{eqnarray}
which are still comprehensible and can be interpreted. Similarly, for the velocity limit we have
\begin{equation}\label{VV}
V_{lim}=\frac{2(1-\omega_H d)}{1+(1-\omega_H d)^2} +{\cal O}(\xi^2)\,.
\end{equation}
The principal novelty here is that $\sin\alpha_{obs}$ depends on the relative velocity $V$ as in Eq. (\ref{finA})  just from the first order of the expansion which may be observationally accessible. In contrast, the expansion (\ref{finZ}) has a first term independent on $\xi$ recovering just the redshift formula we derived recently for a point-like source moving along the ${\bf e}_1$ axis \cite{CDop}. Therefore,  the influence of the black hole dimensions could be observed only when the second term of the  order  ${\cal O}(\xi^2)$ could be measured with a satisfactory accuracy. 

\subsection{Flat limit}
 
The limit of the de Sitter relativity when the de Sitter-Hubble constant $\omega_H$ vanishes is just the usual special relativity. Then all the co-moving frames of the de Sitter relativity become inertial frames in Minkowski space-time without affecting the black hole geometry in its proper frame. Now a remote observer sees a photon emitted in $S$ as having an apparent  rectilinear trajectory with momentum $k$ and energy $E=k$.  Then all the measured quantities can be obtained from the de Sitter ones in the limit $\omega_H\to 0$.   In this limit our principal parameters becomes
\begin{equation}
r_S\to\hat r_S=3\sqrt{3}M\,,\quad \xi\to \hat\xi=\frac{3\sqrt{3}M}{d}
\end{equation}
while the other quantities have the limits 
\begin{eqnarray}
\sin\hat\alpha&=&\lim_{\omega_H\to 0}\sin\alpha= (1-V)\, \hat\xi\, \Delta\,,\label{sasp0}\\
\sin\hat\alpha_{obs}&=&\lim_{\omega_H\to 0}\sin\alpha_{obs}= \sqrt{1-V^2} \, \hat\xi\, \,,\label{sasp}\\
\frac{1}{1+\hat z}&=&\lim_{\omega_H\to 0}\frac{1}{1+z}=\sqrt{\frac{1-V}{1+V}}\,\Delta \,,\label{rssr}
\end{eqnarray}
where
\begin{eqnarray}
\Delta&=&\frac{1+V}{1+V\cos\hat\alpha_{obs} }=\frac{1+V}{1+V\sqrt{1-\hat\xi^2(1-V^2)}}\nonumber\\
&=&1+V(1-V)\,\frac{\hat\xi^2}{2}+{\cal O}(\hat\xi^4)\,.
\end{eqnarray}
As expected, hereby we recover the usual aberration formulas
\begin{equation}
\sin\hat\alpha=\frac{\sqrt{1-V^2}\sin\hat\alpha_{obs}}{1+V\cos\hat\alpha_{obs}}\Leftrightarrow \sin\hat\alpha_{obs}=\frac{\sqrt{1-V^2}\sin\hat\alpha}{1-V\cos\hat\alpha}\,.
\end{equation}
In addition, we obtain the limit velocity  
\begin{equation}\label{vasp}
\hat V_{\lim}=\lim_{\omega_H\to 0}V_{lim}=\frac{1}{\cos\hat\alpha}\ge 1\,,
\end{equation}
does not make sense exceeding the speed of light when $\hat\alpha\to 0$. Thus the limitation of velocities disappears together with the de Sitter event horizon. 

For interpreting these results a good choice is $\Delta\simeq 1$ as the parameter $\xi$ remains very small. Then Eq. (\ref{rssr}) is just the redshift due to the Doppler effect in special relativity. Moreover, we may convince ourselves that all the limits derived above are just the results that may be obtained by applying our method in special relativity, as presented briefly in the App. B. 

\section{Concluding remarks} 

We studied how a co-moving observer measures simultaneously the shadow and redshift of a Scwarzschild black hole freely falling  in the de Sitter expanding universe. 
For this purpose we used a new algebraic method offered by our de Sitter relativity that offers suitable isometries transforming the conserved quantities of the emitted photon  into those recorded by a remote co-moving observer. In this manner we obtained the closed formula (\ref{finA}) of the shadow depending on the peculiar velocity and the corrections to the new redshift formula (\ref{finZ}) that can be calculated on computer by using the code \cite{Comp}.

Another advantage of our method is that this is somewhat independent on the coordinates which are involved only in imposing the initial conditions. For example, the choice of the co-moving frames with Painlev\' e coordinates simplifies the calculations  since then we use the translation  (\ref{trap}) which does not  affect the time. In contrast, in static coordinates, defined by Eq.  (\ref{static}), the same translation gives the transformation (\ref{stattr}) which affects the time such that it is more difficult to synchronise the clocks by setting common initial conditions when $V\not=0$. However, this is not a real impediment as long as we know how the coordinates transform among themselves.

This relative independence on coordinates can be tested in the case of $V=0$ applying our method to a static black hole. Our preliminary calculations indicate that  the shadow of the static black hole is given by  Eq. (\ref{fina})  just as in the case of the co-moving frames with Painlev\' e coordinates. This stability comes from the fact that in both these cases we start with the same conserved quantities (\ref{Q1}) - ({\ref{Q2}) transformed by the same translation (\ref{tr}) whose parameter $d$ is the physical distance  between black hole and observer. 

Under such circumstances we may compare how  the algebraic and geometric methods work in determining the black hole shadow at least in the case of  $V=0$.  The shadow formula derived in  Ref. \cite{SS1}  by using the geometric method can be written in our notation as  $\sin \alpha_{obs}= \frac{r_S}{d'}\sqrt{f(d')}$ where $d'$ is now the radial coordinate of the fixed observer. Thus we see that in our approach the factor $\sqrt{f(d')}$ is missing.  This means that between these two methods there are some minor differences that may come from the approximation of remote observers on which the algebraic method is based and from the fact that in the static frame the radial coordinate $d'$ does not coincide with the physical distance $d$.  However, now it is premature to say more about this relationship before analysing  many examples in de Sitter relativity. 

In other respects, we must specify that the study of the conserved quantities is not enough for understanding the entire information carried out by the light emitted by moving black holes. There are important observable quantities  resulted from the coordinate transformations under isometries as, for example, the photon propagation time or the real distance between observer and black hole  at the time when the photon is measured. In Ref. \cite{CDop} we derived such quantities in the longitudinal case of a point-like source moving along the direction observer-black hole. Therefore,  when we apply the algebraic method the coordinate transformations under isometries or other geometric tools may complete our investigation. 

We hope that the algebraic method proposed here will improve the general geometric approach  for getting over the difficulties in analysing the light emitted by various cosmic objects moving in the de Sitter expanding universe.

\appendix
\section{Translations}
\setcounter{equation}{0} \renewcommand{\theequation}
{A.\arabic{equation}}

The space translations of the $SO(1,4)$ group are less used in applications such that it is worth reviewing briefly their action in different coordinates.  A space translation of parameters ${\bf a}=(a^1,a^2,a^3)$ is the isometry $x'\to x=\phi_{\frak g({\bf a})}(x')$ generated by the $SO(1,4)$ transformation \cite{Nach,CdSR1},  
\begin{equation}\label{tr1}
{\frak g}({\bf a})=\left(
\begin{array}{ccccc}
1+\frac{1}{2}\,\omega_H^2 {\bf a}^2&\omega_H a^1&\omega_H a^2&\omega_H a^3&\frac{1}{2}\,\omega_H^2 {\bf a}^2\\
\omega_H a^1&1&0&0&\omega_H a^1\\
\omega_H a^2&0&1&0&\omega_H a^2\\
\omega_H a^3&0&0&1&\omega_H a^3\\
-\frac{1}{2}\,\omega_H^2 {\bf a}^2&-\omega_H a^1&-\omega_H a^2&-\omega_H a^3&1-\frac{1}{2}\,\omega_H^2 {\bf a}^2
\end{array}\right)\,,
\end{equation}
according to Eq. (\ref{iso}) where now ${\frak g}={\frak g}({\bf a})$. Solving this equation in the co-moving frames with Painlev\' e coordinates where the $z$-functions have the form (\ref{Zx}) we find the transformation
\begin{equation}\label{trap}
\begin{array}{lll}
t&=&t'\\
{\bf x}&=&{\bf x}'+{\bf a}\,e^{\omega_H t}
\end{array} \,,
\end{equation}
which does not affect the time but is not static. The only genuine static translations transform the conformal coordinates 
\begin{equation}
\begin{array}{lll}
t_c&=&-\frac{1}{\omega_H}e^{-\omega_H t}\\
{\bf x}_c&=&{\bf x}e^{-\omega_H t}
\end{array} 
~~ \to ~~
\begin{array}{lll}
t_c&=&t_c'\\
{\bf x}_c&=&{\bf x}_c'+{\bf a}
\end{array} \,.
\end{equation}

The the static coordinates $(t_s, {\bf x})$  where
\begin{equation}\label{static}
t_s=t-2\ln \chi(r)\,, \quad  \chi(r)=\sqrt{1-\omega_H^2  r^2}\,,
\end{equation}
with $r=|{\bf x}|$, as defined by Eq. (\ref{subs1}), can be introduced by the functions  
\begin{eqnarray}
z^0(t_s,{\bf x})&=&\frac{1}{\omega_H}\,\chi(r) \sinh \omega_H t_s\,,
\nonumber\\
z^i(t_s,{\bf x})&=&x^i \,, \label{Zxs1}\\
z^4(t_s,{\bf x})&=&\frac{1}{\omega_H}\,\chi(r) \cosh \omega_H t_s\,.
\nonumber 
\end{eqnarray}
Solving again  Eq. (\ref{iso}) for these functions and the transformation (\ref{tr1}) we find the transformation rules under translations, 
\begin{eqnarray}
t_s&=&\frac{1}{2\omega_H}\,\ln \left( \frac{\chi(r')}{\chi(r') e^{-2\omega_H t_s'}-2\omega_H^2\,{\bf a}\cdot{\bf x}'e^{-\omega_H t_s'}-\omega_H^2 {\bf a}^2\chi(r') } \right)\,,\nonumber\\
{\bf x}&=&{\bf x}'+{\bf a}\chi(r')\, e^{\omega_H t_s'}\,,\label{stattr}
\end{eqnarray}
we present here for the first time.

\section{Moving black holes in special relativity}
\setcounter{equation}{0} \renewcommand{\theequation}
{B.\arabic{equation}}

Let us consider now a Schwarzschild black hole embedded in the Minkowski flat space-time where the remote observers stay in inertial frames. We apply our algebraic method assuming that at the initial time $t=0$ the black hole origin $O_{BH}$ is translated with $d$ and moves with the relative velocity ${\bf V}=(V,0,0)$ with respect to $O$.  At the same time, the apparent source $S$ of coordinates 
\begin{equation}
x'=(0,-\hat r_S \sin\hat\alpha,-\hat r_S \cos\hat\alpha,0)^T
\end{equation}  
emits a photon of energy-momentum 
\begin{equation}
p'=(k, -k\cos\hat\alpha, -k\sin\hat\alpha,0)^T\,,
\end{equation}
which has to be observed in the fixed origin $O$.  

The  black hole and observer frames, $O_{BH}$ and $O$, are related through the isometry  $\Lambda=L({\bf V})T(d)$ formed by the translation $T(d):{x'}^1\to{x}^1= {x'}^1+d$  followed by a Lorentz transformation $L({\bf V})$, which is just the four-dimensional restriction of the matrix (\ref{Lor}). Performing this isometry  we obtain the four-vectors $x=\Lambda x'$ and $p=\Lambda p'$ whose components give the energy and angular momentum observed in $O$ as
\begin{eqnarray}
E&=&p^0=\frac{k(1-V\cos\hat\alpha)}{\sqrt{1-V^2}}\,,\label{Eap}\\
L_3&=&x^1p^2-x^2p^1=\frac{kd\left[\xi(1-V\cos\hat\alpha)-\sin\hat\alpha \right]}{\sqrt{1-V^2}}\,.
\end{eqnarray}
The condition $L_3=0$ of the photon passing through $O$  gives angle $\alpha$ in the black hole frame,
\begin{equation}
\tan\frac{\hat\alpha}{2}=\frac{1-\sqrt{1-\hat\xi^2(1-V^2)}}{\hat\xi(1+V)}\,.
\end{equation} 
In observer's frame the black hole shadow is seen under the angle $\hat\alpha_{obs}$ defined as
\begin{equation}\label{aaa}
\sin\hat\alpha_{obs}=\frac{|p^2|}{p}\,,\quad p=\sqrt{{p^1}^2+{p^2}^2}\,.
\end{equation} 
Finally, calculating $\sin\hat\alpha$ and $\cos\hat\alpha$ and substituting these values in Eqs. (\ref{Eap}) and (\ref{aaa}) we verify that the limits (\ref{sasp0}-\ref{vasp}) are correct.


\begin{thebibliography}{}
\bibitem{Comp}
I. I. Cot\u aescu,  {\em Maple code BH01} (2020) download here \href{https://physics.uvt.ro/~cota/CCFT/codes}{BH01}

\bibitem{L1}
G. E. Lema\^ itre, {\em Ann. Soc. Sci. de Bruxelles} {\bf 47A} (1927) 49.

\bibitem{L2}
G. E. Lema\^ itre, {\em MNRAS} {\bf 91} (1931) 483.

\bibitem{Hubb}
E. Hubble, {\em Proc. Nat. Acad. Sci.} {\bf 15} (1929) 168.

\bibitem{SW}
S. Weinberg, {\em Gravitation and Cosmology: Principles and Applications of the General Theory of relativity} (J. Wiley \& Sons, New York 1972). 

\bibitem{H0}
E. R. Harrison, {\em  Cosmology: The Science of the Universe} (New York: Cambridge Univ. Press, 1981).

\bibitem{H}
E. Harrison, {\em  Astrophys. J.} {\bf 403} (1993) 28.

\bibitem{LL}
L. D. Landau and E. M. Lifshitz, {\em The classical theory of fields} (Elsevier Sci. Inc. NY. 1975).

\bibitem{CdSR1}
I. I. Cot\u aescu, {\em Eur. Phys. J. C} {\bf 77} (2017) 485.

\bibitem{CdSR2}
I. I. Cot\u aescu, {\em Eur. Phys. J. C} {\bf 78}  (2018) 95.

\bibitem{CDop}
I. I. Cot\u aescu, arXiv: 2007.13590 (accepted for publication in MPLA).

\bibitem{BH1}
K. Akiyama, et al., {\em Astrophys. J.} {\bf 875}(1) (2019) L1.  
\bibitem{BH2}
K. Akiyama, et al., {\em Astrophys. J.} {\bf 875}(1) (2019) L6. 

\bibitem{Sy} 
J. L. Synge, {\em Mon. Not. R. Astron. Soc.} {\bf 131} (1966) 463.

\bibitem{SS1}% McVittie
V. Perlick, O. Yu. Tsupko, G. S. Bisnovatyi-Kogan, {\em Phys. Rev. D} {\bf 97} (2018) 104062.

\bibitem{SS2}
G. S. Bisnovatyi-Kogan and O. Yu. Tsupko,  {\em Phys. Rev. D} {\bf 98} (2018) 084020. 

\bibitem{SS01}
J. T. Firouzjaee and A. Allahyari, {\em Eur. Phys. J. C} {\bf 79} (2019) 1140.

\bibitem{SS02}
Z. Chang and Q.-H. Zhu, {\em JCAP} {\bf 06} (2020) 055.

\bibitem{SS03}
S. Vagnozzi, C. Bambi, and L. Visinelli, {\em Class. Quantum Grav.} {\bf 37} (2020) 087001. 

\bibitem{SS3}
O. Yu. Tsupko G. S. Bisnovatyi-Kogan, {\em  Int.  J.  Mod. Phys. D} {\bf 29} (2020) 2050062. 

\bibitem{K1}
J. M. Bardeen, {\em Proceedings, Ecole d'Et de Physique Thorique: Les Astres Occlus.} (Les Houches, France, 1973).
\bibitem{K2}
H. Falcke, F. Melia, and E. Agol, {\em Astrophys. J. Lett.} {\bf 528} (1999) L13.

\bibitem{KdS1}
A. Grenzebach, V. Perlick, and C. L\" ammerzahl, {\em Phys. Rev. D} {\bf 89} (2014) 124004.
\bibitem{KdS2}
Z. Stuchlik, D. Charbul\' ak, and J. Schee, {\em Eur. Phys. J. C} {\bf 78} (2018) 180.
\bibitem{KdS3}
P.-C. Li, M. Guo, and B. Chen, {\em Phys. Rev. D} {\bf 101} (2020) 084041. 
\bibitem{KdS4}
Z. Chang and Q.-H. Zhu, {\em Phys. Rev. D} {\bf 101} (2020) 084029.

\bibitem{M1}
F. Atamurotov, A. Abdujabbarov, and B. Ahmedov, {\em Phys. Rev. D} {\bf 88} (2013) 064004.
\bibitem{M2}
Z. Li and C. Bambi, {\em J. Cosmol. Astropart. Phys.} {\bf 2014}  (01) 041.
\bibitem{M3}
A. Grenzebach, V. Perlick, and C. L\" ammerzahl, {\em Int. J. Mod. Phys. D} {\bf 24} (2015) 1542024.
\bibitem{M4}
P. V. P. Cunha, C. A. R. Herdeiro, E. Radu, and H. F. Runarsson, {\em Phys. Rev. Lett.} {\bf 115} (2015) 211102.
\bibitem{M5}
 A. Abdujabbarov, M. Amir, B. Ahmedov, and S. G. Ghosh, {\em Phys. Rev. D} {\bf 93} (2016) 104004.
\bibitem{M6}
 C. Bambi, K. Freese, S. Vagnozzi, and L. Visinelli, {\em Phys. Rev. D} {\bf 100} (2019) 044057. 
 \bibitem{M7}
S. Vagnozzi and L. Visinelli, {\bf Phys. Rev. D} {\bf 100} (2019) 024020.
\bibitem{M8}
K. Jusu, M. Jamil, P. Salucci, T. Zhu, and S. Haroon, {\em Phys. Rev. D} {\bf 100} (2019) 044012.
\bibitem{M9}
 P. V. P. Cunha, C. A. R. Herdeiro, and E. Radu, {\em Universe} {\bf 5} (2019) 220. 
\bibitem{M10}
S. Haroon, M. Jamil, K. Jusu, K. Lin, and R. B. Mann, {\em Phys. Rev. D 99} (2019)  044015.
\bibitem{M11}
S.-W. Wei, Y.-X. Liu, and R. B. Mann, {\em Phys. Rev. D} {\bf 99} (2019) 041303.
\bibitem{M12}
N. Bar, K. Blum, T. Lacroix, and P. Panci, {\em J. Cosmol. Astropart. Phys.} {\bf 2019} (07) 045. 
\bibitem{M13}
H. Davoudiasl and P. B. Denton, {\em Phys. Rev. Lett.} {\bf 123}  (2019)  021102. 
\bibitem{M14}
I. Banerjee, S. Chakraborty, and S. SenGupta, {\em Phys. Rev. D} {\bf 101} (2020) 041301.
\bibitem{M15}
R. Roy and U. A. Yajnik, {\em Phys. Lett. B} {\bf 803} (2020) 135284. 
\bibitem{M16}
C. Li, S.-F. Yan, L. Xue, X. Ren, Y.-F. Cai, D. A. Easson, Y.-F. Yuan, and H. Zhao, {\em  Phys. Rev. Research} {\bf 2} (2020) 023164. 

\bibitem{Pan}
P. Painlev\' e, {\em  C. R. Acad. Sci.} (Paris) {\bf 173} (1921) 677.

\bibitem{Kot}
F. Kottler, {\em Ann. Phys.} (Berlin) {\bf 361} (1918) 401.

\bibitem{CGRG}
I. I. Cot\u aescu, {\em GRG} {\bf 43} (2011) 1639.

\bibitem{Gib}
G. W. Gibbons, C. M. Warnick, and M. C. Werner, {\em Class. Quantum Grav.} {\bf 25} (2008) 245009.

\bibitem{El1}
T. Treu, R.S. Ellis, {\em Contemporary Physics} {\bf 56}(1) (2015) 17.

\bibitem{Ed}
A. S. Eddington,   {\em The mathematical theory of relativity} 
(Cambridge University Press 1923).

\bibitem{D1}
 C. Darwin, Proc. Roy. Soc. {\bf 249} (1958) 180.
 
\bibitem{D2}
 C. Darwin, Proc. Roy. Soc. {\bf 263} (1961) 39.
 
 \bibitem{V1}
C. M. Claudel, K.S. Virbhadra and G. F. R. Ellis, {\em J. Math. Phys.} \textbf{42}  (2001) 818.
 
\bibitem{V2}
K. S. Virbhadra and G. F. R. Ellis,
%``Schwarzschild black hole lensing,''
{\em Phys. Rev. D} \textbf{62}, (2000) 084003. 


\bibitem{V3}
K. S. Virbhadra and G. F. R. Ellis,
%``Gravitational lensing by naked singularities,''
{\em Phys. Rev. D} \textbf{65} (2002) 103004. 


\bibitem{V4}
K. S. Virbhadra, D. Narasimha and S. M. Chitre,
%``Role of the scalar field in gravitational lensing,''
{\em Astron. Astrophys.} \textbf{337} (1998) 1.

\bibitem{V5}
K. S. Virbhadra and C. R. Keeton,
%``Time delay and magnification centroid due to gravitational lensing by black holes and naked singularities,''
{\em Phys. Rev. D} \textbf{77} (2008) 124014. 

\bibitem{V6}
K. S. Virbhadra,
%``Relativistic images of Schwarzschild black hole lensing,''
{\em Phys. Rev. D} \textbf{79} (2009) 083004. 
 
 
\bibitem{CdSG}%geodesices
I. I. Cot\u aescu, {\em Mod. Phys. Lett.  A} {\bf 32}  (2017) 1750223. 

\bibitem{Nach}
O. Nachtmann, {\em Commun. Math. Phys.} {\bf 6} (1967) 1.




\end{thebibliography}
\end{document}